\documentclass[12pt]{article}
\usepackage{graphicx}% Include figure files
\usepackage{dcolumn}% Align table columns on decimal point
\usepackage{bm}% bold math

\def\d{\mbox{\rm d}}

\def\half{\mbox{$\frac{1}{2}$}}

\title{Jacobi's Last Multiplier and Lagrangians for Multidimensional Systems}

\author{MC Nucci\footnote{Email: nucci@unipg.it} $\;$  and PGL Leach\footnote{permanent address:  School of Mathematical Sciences, Westville Campus,
University of KwaZulu-Natal, Private Bag X54001 Durban 4000,
Republic of South Africa. Email: leachp@ukzn.ac.za, leach@math.aegean.gr}\\[0.21cm]
 Dipartimento di Matematica e Informatica\\ Universit\`a di Perugia, 06123
 Perugia, Italy}

\begin{document}
\maketitle

\begin{abstract}
We demonstrate that the formalism for the calculation of the Jacobi last multiplier for a one-degree-of-freedom system extends naturally to systems of more than one degree of freedom thereby extending results of Whittaker dating from more than a century ago and Rao ({\it Proc Ben Math Soc} {\bf 2} (1940) 53-59) dating from almost seventy years ago.  We illustrate the theory with an application taken from the theory of coupled oscillators.  We indicate how many Lagrangians can be obtained for such a system.
\end{abstract}



\section{\label{sec:level1}Introduction}

Jacobi's Last Multiplier is a solution of the linear partial
differential
 equation \cite {Jacobi 44 b, Jacobi 45 a, Jacobi 86 a, Whittaker 44 a},
\begin {equation}
\sum_{i = 1} ^N \frac {\partial (Ma_i)} {\partial x_i} = 0,
\label{1.0}
\end {equation}
where $\sum_{i = 1} ^N a_i\partial_{x_i} $ is the vector field of
the set of first-order ordinary differential equations for the $N
$ dependent variables $x_i $, namely
\begin {equation}
Af = \sum_{i = 1} ^n a_i(x_1,\dots,x_n)\frac {\partial f}
{\partial x_i} = 0 \label {12.1}
\end {equation}
or its equivalent associated Lagrange's system
\begin {equation}
\frac {\d x_1} {a_1} = \frac {\d x_2} {a_2} = \ldots = \frac {\d
x_n} {a_n}.\label {12.2}
\end {equation}
The ratio of any two multipliers is a first integral of the system
of first-order differential equations and in the case that this
system is derived from the Lagrangian of a one-degree-of-freedom
system one has that \cite {Jacobi 86 a, Whittaker 44 a}
\begin{equation}
\frac{\partial^2L}{\partial \dot{q}^2} = M. \label{1.2}
\end{equation}
Consequently a knowledge of the multipliers of a system enables
one to construct a number of Lagrangians of that system.

\strut\hfill

We recall that Lie's method \cite {Lie 74 a,Lie 67 a} for the
calculation of the Jacobi Last Multiplier is firstly to find the
value of
\begin{equation}
\Delta = \mbox{\rm det}\left[\begin{array}{c} e_{ij}\\ s_{ij}
\end{array} \right], \label{1.1}
\end{equation}
in which the matrix is square with the elements $e_{ij}$ being the
vector field of the set of first-order differential equations by
which the system is described and the elements, $s_{ij}$, being
the coefficient functions of the number of symmetries of the given
system necessary to make the matrix square. If $\Delta$ is not
zero, the corresponding multiplier is $M=\Delta^{-1}$. There is an
obvious corollary to the results of Jacobi mentioned above. In the
case that there exists a constant multiplier, the determinant is a
first integral.  This result is potentially very useful in the
search for first integrals of systems of ordinary differential
equations.

\strut\hfill

The relationship between the Jacobi Last Multiplier and the
Lagrangian for a one-degree-of-freedom system is perhaps not
widely known although it is certainly not unknown as can be seen
from the bibliography in \cite {Nucci 05 b}.  Given a knowledge of
a multiplier, (\ref {1.1}) gives a simple recipe for the
generation of a Lagrangian.  The only possible difficulty is the
performance of the double quadrature.  In this paper we consider
that the vector fields of the system of equations and symmetries
are known and that we seek the multiplier.  From another direction
one could know the multiplier and all but one of the symmetries.
From (\ref{1.1}) the remaining symmetry can be determined
\cite{Nucci 02 a}. A knowledge of the multipliers of a system
enables one to construct a number of Lagrangians of that system.
Considering the dual nature of the Jacobi Last Multiplier as
providing a means to determine both Lagrangians and integrals one
is surprised that it has not attracted more attention over the
more than one and a half centuries since its introduction.  The
bibliography of \cite {Nucci 05 b} gives a fair indication of its
significant applications in the past to which we may add
\cite{Cosgrove 00 a}, \cite{Cosgrove 06 a}, \cite{Borisov 05 a},
\cite{Nucci 06 a}, \cite{Nucci 06 b}, \cite{Nucci 07 a},
\cite{Nucci 07 b}.

\strut\hfill

\section{\label{sec:level2}Theoretical Development}

In Rao \cite {Rao 40 a} there is a development of the work of Whittaker \cite {Whittaker 44 a} [pp 276-286] on the theory of the Last Multiplier of Jacobi for multidimensional systems.  The discussion by Whittaker is limited to dynamical systems without a potential.  The work of Rao is restricted to dynamical systems with a Lagrangian which is a quadratic form in the generalised velocities.  It does provide a clearer path to an understanding of the methodology than that provided by Whittaker.  In particular Rao gives specific formul\ae\ for the equations to be satisfied by the various multipliers.  We see in the sequel that a considerable simplification occurs, but we pay due acknowledgement to the inspiration afforded by that paper.  Although the theory which we develop here is applicable to dynamical systems of all dimensions, we confine our attention to two dimensions to keep the mathematics as simple as possible so that the essential ideas are quite evident.  We assume that we have a Lagrangian, $L (t,q_1,q_2,\dot {q}_1,\dot {q}_2) $, in a standard notation.  We write the simplified form of the corresponding Euler-Lagrange equations as
\begin {eqnarray}
& &\ddot {q}_1 = f_1 (t,q_1,q_2) \label {2.1}, \\
& &\ddot {q}_2 = f_2 (t,q_1,q_2) \label {2.2},
\end {eqnarray}
where we note that the generalised force is independent of the generalised velocities.  The corresponding vector field is
\begin {equation}
X_1 = \partial_t+\dot {q}_1\partial_{q_1} +\dot
{q}_2\partial_{q_2}+f_1\partial_{\dot {q}_1} +f_2\partial_{\dot
{q}_2}.  \label {2.0}
\end {equation}
The actual Euler-Lagrange equations are
\begin {eqnarray}
& &\displaystyle {\frac {\partial ^ 2 L} {\partial\dot {q}_1\partial t} + \frac {\partial ^ 2L} {\partial\dot {q}_1\partial q_1}\dot {q}_1 + \frac {\partial ^ 2L} {\partial\dot {q}_1\partial q_2}\dot {q}_2 + \frac {\partial ^ 2L} {\partial \dot {q}_1 ^ 2}f_1 + \frac {\partial ^ 2L} {\partial\dot {q}_1\partial\dot {q}_2}f_2 - \frac {\partial L} {\partial q_1}} = 0 \label {2.3} \\
& &\displaystyle {\frac {\partial ^ 2 L} {\partial\dot {q}_2\partial t} + \frac {\partial ^ 2L} {\partial\dot {q}_2\partial q_1}\dot {q}_1 + \frac {\partial ^ 2L} {\partial\dot {q}_2\partial q_2}\dot {q}_2 + \frac {\partial ^ 2L} {\partial \dot {q}_1\partial \dot {q}_2}f_1 + \frac {\partial ^ 2L} {\partial\dot {q}_2 ^ 2}f_2 - \frac {\partial L} {\partial q_2}} = 0, \label {2.4}
\end {eqnarray}
in which we have replaced $\ddot {q}_1 $ and $\ddot {q}_2 $ by the right sides of the equations of motion above.

\strut\hfill

We follow Rao \cite {Rao 40 a} in defining the connection between
the last multiplier and the Lagrangian as
\begin {equation}
M_{ij} = \frac {\partial ^ 2 L } {\partial\dot {q}_i\partial\dot
{q}_j},\,\,i,j = 1,\, 2, \label {2.5}
\end {equation}
{\it ie}, we assume the same relationship as for the
one-degree-of-freedom case so that, once the multiplier has been
calculated, the Lagrangian follows by a double quadrature.

\strut\hfill

We differentiate (\ref {2.3}) and (\ref {2.4}) once with respect
to both $\dot {q}_1 $ and $\dot {q}_2 $.  We illustrate the
calculation in the case of the differentiation of (\ref {2.3})
with respect to $\dot {q}_2 $.  We obtain
\begin {eqnarray}
& &\displaystyle {\frac {\partial ^ 3 L} {\partial\dot {q}_1\partial\dot {q}_2\partial t} + \frac {\partial ^ 3L} {\partial\dot {q}_1\partial\dot {q}_2\partial q_1}\dot {q}_1 + \frac {\partial ^ 3L} {\partial\dot {q}_1\partial\dot {q}_2\partial q_2}\dot {q}_2} \nonumber \\
& &\quad +\displaystyle { \frac {\partial ^ 2L} {\partial\dot {q}_1\partial q_2}+ \frac {\partial ^ 3L} {\partial \dot {q}_1 ^ 2\partial\dot {q}_2}f_1
+ \frac {\partial ^ 3L} {\partial\dot {q}_1\partial\dot {q}_2 ^ 2}f_2 - \frac {\partial ^ 2L} {\partial\dot {q}_2\partial q_1}} = 0, \label {2.6}
\end {eqnarray}
in which we have used the independence of $f_1 $ and $f_2 $ from $\dot {q}_2 $ (in this instance).  We use the definition, (\ref {2.5}), for $M_{12} $.  Then (\ref {2.6}) becomes
\begin {eqnarray}
& &\displaystyle {\frac {\partial} {\partial t}\left (M_{12}\right) +\frac {\partial} {\partial q_1}\left (M_{12}\dot {q}_1\right) +\frac {\partial} {\partial q_2}\left (M_{12}\dot {q}_2\right) + \frac {\partial} {\partial\dot {q}_1}\left (M_{12}f_1\right)} \nonumber \\
& &\quad +\displaystyle { \frac {\partial} {\partial\dot {q}_2}\left (M_{12}f_2\right) +\frac {\partial ^ 2L} {\partial\dot {q}_1\partial q_2} -
\frac {\partial ^ 2L} {\partial\dot {q}_2\partial q_1}} = 0. \label {2.65}
\end {eqnarray}
In the cases of $M_{11} $ and $M_{22} $ the ultimate and penultimate terms in the equations corresponding to (\ref {2.65}) cancel.  The other equation for $M_{12} $ has the subscripts reversed so that, when the two equations are added, the terms vanish.  Consequently each multiplier is a solution of the equation
\begin {equation}
\frac {\partial} {\partial t}\left (M\right) +\frac {\partial} {\partial q_1}\left (M\dot {q}_1\right) +\frac {\partial} {\partial q_2}\left (M\dot {q}_2\right) + \frac {\partial} {\partial\dot {q}_1}\left (Mf_1\right) +
\frac {\partial} {\partial\dot {q}_2}\left (Mf_2\right) = 0. \label {2.7}
\end {equation}
Allowing for the variations in notation (\ref {2.7}) is the same
equation as (\ref {1.0}), {\it ie}, the number of degrees of
freedom of the system only affects the number of terms in the
equation.

\strut\hfill

\section {\label{sec:level3}Example: A two-dimensional coupled oscillator}

\subsection {Equivalent Lagrangians}

The collinear motion of two particles of equal mass connected to
each other by a spring and to fixed points by two other springs of
equal spring constants is modelled by the fourth-order system\footnote{A simplified version of this system can be found in
\cite{China}[pp 479-481]}.
\begin {eqnarray}
& &\ddot {q}_1 = -\Omega_1 ^ 2q_1- \Omega_2 ^ 2 (q_1-q_2) \label {3.1} \\
& &\ddot {q}_2 = \Omega_2^2 (q_1-q_2) -\Omega_1 ^ 2q_2. \label {3.2}
\end {eqnarray}
It is evident that a solution of (\ref {2.7}) is simply $M = constant $.  We take the constant to be unity for $M_{11} $ and $M_{22} $ and zero for $M_{12} $.  Then the corresponding Lagrangian is
\begin {equation}
L = \half\left (\dot {q}_1 ^ 2+\dot {q}_2 ^ 2\right) +h_1 (t,q_1,q_2)\dot {q}_1+h_2 (t,q_1,q_2)\dot {q}_2+h_3 (t,q_1,q_2), \label {3.3}
\end {equation}
where the three functions, $h_i $, $i = 1,\, 3 $, are functions of integration.  When we substitute $L $, (\ref {3.3}), into the general Euler-Lagrange equations, we require that we obtain equations (\ref {3.1}) and (\ref {3.2}).  This requirement imposes the constraints
\begin {eqnarray}
\displaystyle {\frac {\partial h_1} {\partial q_2} -\frac {\partial h_2} {\partial q_1}} = 0 & & \label {3.4} \\
\displaystyle {\frac {\partial h_1} {\partial t} -\frac {\partial h_3} {\partial q_1}}  -\Omega_1 ^ 2q_1- \Omega_2 ^ 2 (q_1-q_2) = 0 & & \label {3.5}  \\
\displaystyle {\frac {\partial h_2} {\partial t} -\frac {\partial h_3} {\partial q_2}}  + \Omega_2^2 (q_1-q_2) -\Omega_1 ^ 2q_2 = 0, & & \label {3.6}
\end {eqnarray}
where the constraint (\ref {3.4}) is common to both equations.  The constraint (\ref {3.4}) is the two-dimensional version of the condition for the functions to be the divergence of some arbitrary function, $g (t,q_1,q_2) $, {\it ie}
\begin {equation}
h_1 = \frac {\partial g} {\partial q_1} \quad\mbox {\rm and}\quad h_2 = \frac {\partial g} {\partial q_2}. \label {3.7}
\end {equation}
When (\ref {3.7}) is substituted into (\ref {3.5}) and (\ref {3.6}), one obtains
\begin {eqnarray}
\displaystyle {\frac {\partial ^ 2g} {\partial t\partial q_1} -\frac {\partial h_3} {\partial q_1}} -\Omega_1 ^ 2q_1- \Omega_2 ^ 2 (q_1-q_2) = 0 & & \label {3.8}  \\
\displaystyle {\frac {\partial ^ 2g} {\partial t\partial q_2}
-\frac {\partial h_3} {\partial q_2}}  + \Omega_2^2 (q_1-q_2)
-\Omega_1 ^ 2q_2 = 0, & & \label {3.9}
\end {eqnarray}
which can be consistently integrated to give
\begin {equation}
h_3 = \frac {\partial g} {\partial t} -\half\left [\left (\Omega_1 ^ 2+\Omega_2 ^ 2\right)\left (q_1 ^ 2+q_2 ^ 2\right) - 2\Omega_2 ^ 2q_1q_2\right] + s (t), \label {3.10}
\end {equation}
where $s (t) $ is an arbitrary function which may as well be set at zero, since it makes no contribution to the Euler-Lagrange equation, or included in the other arbitrary function $g (t,q_1,q_2) $ which also makes no contribution to the Euler-Lagrange equation.

\strut\hfill

Eventually we have arrived at the Lagrangian
\begin {equation}
L = \half\left (\dot {q}_1 ^ 2+\dot {q}_2 ^ 2\right) + \dot {g} -\half\left [\left (\Omega_1 ^ 2+\Omega_2 ^ 2\right)\left (q_1 ^ 2+q_2 ^ 2\right) - 2\Omega_2 ^ 2q_1q_2\right], \label {3.11}
\end {equation}
{\it ie} the arbitrariness in the Lagrangian can be expressed as a total time derivative.  Such a Lagrangian has been termed `gauge variant' \cite {Levy-Leblond 71 a} and is notable in that the presence of the arbitrary function has no effect upon the number of Noether point symmetries \cite {Nucci 07 a}.  In this respect it could be regarded as part of the boundary term in the way Noether put it in her formulation of her theorem \cite {Noether 18 a}.  The class of Lagrangians described by (\ref {3.11}) is an equivalence class.

\strut\hfill

\subsection {A Plethora of Multipliers}

Jacobi proposed to obtain the last multiplier from the solution of the first-order linear partial differential equation, (\ref {2.7}).  In 1874 Lie \cite {Lie 74 a} showed that one could use point symmetries \footnote {At that time the only symmetries considered were point symmetries. With the extension of the concept of infinitesimal transformations generated by symmetries depending upon derivatives and integrals there has been a natural extension of Lie's valuable contribution to the subject of the Jacobi Last Multiplier.} to determine last multipliers. Lie's method for the calculation of the Jacobi Last Multiplier is firstly to find the value of the determinant in (\ref {2.7}).  The number of symmetries required depends upon the number of elements in the vector field, which supplies the first row of the matrix, since the matrix must be square.  Here we consider that the vector fields of the system of equations and symmetries are known and that we seek the multiplier.  The Lie point symmetries of system (\ref {3.1}/\ref {3.2}) are
\begin {eqnarray}
& &\Gamma_1 = q_2\partial_{q_1} +q_1\partial_{q_2} +\dot {q}_2\partial_{\dot {q}_1} +\dot {q}_1\partial_{\dot {q}_2} \nonumber \\
& &\Gamma_2 = \exp\left [Q it\right]\left[\left (\partial_{q_1}
-\partial_{q_2}\right)
+iQ \left (\partial_{\dot {q}_1}-\partial_{\dot {q}_2}\right)\right] \nonumber\\
& &\Gamma_3 = \partial_t \nonumber\\
& &\Gamma_4 = q_1\partial_{q_1} +q_2\partial_{q_2} +\dot {q}_1\partial_{\dot {q}_1}
 +\dot {q}_2\partial_{\dot {q}_2} \nonumber\\
& &\Gamma_5 = \exp\left [-Q it\right]\left (\partial_{q_1} -\partial_{q_2}
  - iQ \left (\partial_{\dot {q}_1}-\partial_{\dot {q}_2}\right)\right] \nonumber\\
& &\Gamma_6 = \exp\left[i\Omega_1 t\right]\left[\left (\partial_{q_1}
+\partial_{q_2}\right) + i\Omega_1\left (\partial_{\dot {q}_1}
 +\partial_{\dot {q}_2}\right)\right] \nonumber\\
& &\Gamma_7 = \exp\left[-i\Omega_1 t\right]\left[\left
(\partial_{q_1} +\partial_{q_2}\right) -i\Omega_1\Omega_1 t\left
(\partial_{\dot {q}_1} +\partial_{\dot {q}_2}\right)\right],
\nonumber
\end {eqnarray}
where $Q = \sqrt {\Omega_1^2 + 2\Omega_2^2}$, which is the minimal
number of Lie point symmetries for an autonomous second-order
two-dimensional linear system \cite {Gorringe 88 a}. Despite the
fact that the system possesses only the minimal number of Lie
point symmetries there are thirty-five different determinants to
be evaluated.  Naturally not all of these can be expected to be
nonzero or different, but one can easily imagine that this would
be the case for a substantial proportion since in the case of a
single linear oscillator the number of different multipliers is
fourteen out of a possible twenty-eight determinants \cite {Nucci
07 a}.

\strut\hfill

It is not our intention to display a large number of multipliers.  We simply wish to illustrate two points.  The first is the determination of a Lagrangian for the system (\ref {3.1}/\ref {3.2}) differing from the class represented by (\ref {3.11}) which is of a standard form.  The second is to display a first integral for the system by taking the ratio of two autonomous multipliers.  From the Lagrangian (\ref {3.11}) one can construct an autonomous Hamiltonian which is a first integral.  Thus we have two first integrals and, as if we did not already know it, the system is integrable from the Theorem of Liouville.

\strut\hfill

If we take the symmetries $\Gamma_1 $, $\Gamma_2 $, $\Gamma_3 $ and $\Gamma_5 $, the matrix for which the determinant is to be evaluated is
\begin {equation}
C_{1235} = \left [\begin {array} {lllll}
1 &\dot {q}_1 &\dot {q}_2 &-\Omega_1 ^ 2q_1- \Omega_2 ^ 2 (q_1-q_2) &\Omega_2 (q_1-q_2) -\Omega_1 ^ 2q_2\\
0 &q_2 &q_1 &\dot {q}_2 &\dot {q}_1\\
0 &\exp\left [Q it\right] & -\exp\left [Q it\right] &iQ\exp\left [Q it\right] & -iQ\exp\left [Q it\right]\\
1 & 0 & 0 & 0 & 0\\
0 &\exp\left [-Q it\right]  & -\exp\left [-Q it\right]  &
-iQ\exp\left [-Q it\right] &iQ\exp\left [-Q it\right]
\end {array}\right]. \label {3.12}
\end {equation}
We obtain the
multiplier
\begin {equation}
JLM_{1235} = -\frac {1} {2iQ\left [\left (\dot {q}_1+\dot {q}_2\right) ^ 2+\Omega_1 ^ 2\left (q_1+q_2\right) ^ 2\right]}. \label {3.13}
\end {equation}
Similarly by taking the symmetries $\Gamma_1$, $\Gamma_6$, $\Gamma_3$ and $\Gamma_7$ we obtain
\begin {equation}
JLM_{1637} = \frac {1} {2\Omega_2\left (q_1-q_2\right) ^ 2+\Omega_1\left (q_1-q_2\right) ^ 2+\left (\dot {q}_1-\dot {q}_2\right) ^ 2}. \label {3.14}
\end {equation}
Obviously an integral is
\begin {equation}
I_{1637/1235} = \frac {2iQ\left [\left (\dot {q}_1+\dot {q}_2\right) ^ 2+\Omega_1 ^ 2\left (q_1+q_2\right) ^ 2\right]}{2\Omega_2\left (q_1-q_2\right) ^ 2+\Omega_1\left (q_1-q_2\right) ^ 2+\left (\dot {q}_1-\dot {q}_2\right) ^ 2}. \label {3.15}
\end {equation}

\strut\hfill

We can obtain a Lagrangian from $JLM_{1235} $, say, by integrating twice with respect to $\dot {q}_1 $.  We have
\begin {eqnarray}
&&L_{1235%,\dot {q}_1
} = -\frac {1} {2iQ\Omega_1\left (q_1+q_2\right)}\left (\dot {q}_1+\dot
{q}_2\right)\arctan\left (\frac {\dot {q}_1+\dot {q}_2}
{\Omega_1\left (q_1+q_2\right)}\right) \label {3.16}  \\
&& \qquad +\frac {1} {4iQ}\log\left [\frac {\Omega_1 ^ 2\left (q_1+q_2\right) ^ 2+\left
(\dot {q}_1 +\dot {q}_2 \right)^2} {\Omega_1 ^ 2\left
(q_1+q_2\right) ^ 2}\right] + R_1 (t,q_1,q_2,\dot {q}_2)\dot {q}_1
+R_2 (t,q_1,q_2,\dot {q}_2), \nonumber
\end {eqnarray}
where $R_1 $ and $R_2 $ are functions of integration.  When we apply the Euler-Lagrangian condition to (\ref {3.16}) and substitute the required equations of motion, (\ref {3.1}/\ref {3.2}), we obtain the conditions
\begin {equation}
\frac {i} {2Q (q_1+q_2)} +\left [\Omega_2 ^ 2\left (q_1 -q_2\right) -\Omega_1 ^ 2q_2\right]\frac {\partial R_1} {\partial\dot {q}_2} +\dot {q}_2\frac {\partial R_1} {\partial q_2} -\frac {\partial R_1} {\partial t} -\frac {\partial  R_2} {\partial q_1} = 0  \label {3.17}
\end {equation}
and
\begin {eqnarray}
& &\frac {i} {2Q (q_1+q_2)} -\left [\Omega_1 ^ 2q_1+\Omega_2 ^2\left(q_1 -q_2\right)\right]\displaystyle{\frac {\partial R_1} {\partial\dot {q}_2}} -\dot {q}_1\displaystyle {\frac {\partial R_1} {\partial q_1}} -\displaystyle {\frac {\partial R_2} {\partial q_2}} +\dot {q}_1\left [\Omega_2 ^2\left (q_1 -q_2\right) -\Omega_1 ^ 2q_2\right]\displaystyle {\frac {\partial ^ 2 R_1} {\partial \dot {q}_2 ^ 2}} \nonumber \\
& & +\dot {q}_1\dot {q}_2\displaystyle {\frac {\partial ^ 2 R_1} {\partial q_2 \partial\dot {q}_2}}+\dot {q}_1 ^ 2\displaystyle {\frac {\partial ^ 2 R_1} {\partial q_1 \partial\dot {q}_2}} +\dot {q}_1\displaystyle {\frac {\partial ^ 2 R_1} {\partial t ^ 2}} = 0. \label {3.18}
\end {eqnarray}
This second equation is a quadratic in $\dot {q}_1 $ which is not one of the variables in $R_1 $ and $R_2 $.  The three coefficients give
\begin {eqnarray}
\displaystyle {\frac {\partial ^ 2 R_1} {\partial q_1 \partial\dot {q}_2}} & = & 0 \label {3.19} \\
-\displaystyle {\frac {\partial R_1} {\partial q_1}} +\left [\Omega_2 ^2\left (q_1 -q_2\right) -\Omega_1 ^ 2q_2\right]\displaystyle {\frac {\partial ^ 2 R_1} {\partial \dot {q}_2 ^ 2}} +\dot {q}_2\displaystyle {\frac {\partial ^ 2 R_1} {\partial q_2 \partial\dot {q}_2}}+\displaystyle {\frac {\partial ^ 2 R_1} {\partial t ^ 2}} & = & 0 \label {3.20} \\
\frac {i} {2Q (q_1+q_2)} -\left [\Omega_1 ^ 2q_1+\Omega_2 ^2\left (q_1 -q_2\right)\right]\frac {\partial R_1} {\partial\dot {q}_2}-\displaystyle {\frac {\partial R_2} {\partial q_2}} & = & 0. \label {3.21}
\end {eqnarray}
We note that $R_2 $ appears only in (\ref {3.17}) and (\ref {3.21}) and in a very symmetrical way at that.  Equations (\ref {3.19}) and (\ref {3.20}) are easily satisfied by setting $R_1 $ equal to zero.  It then follows that we can take
\begin {equation}
R_2= \frac {i} {2Q}\log\left (q_1+q_2\right). \label {3.22}
\end {equation}

\strut\hfill

\section {\label{sec:level4}Discussion}

The last multiplier of Jacobi has in the main been sadly
neglected. It is evident from the slight literature \cite
{Whittaker 44 a, Rao 40 a} that its application to systems of more
than one degree of freedom was either ignored or misunderstood. In
\S 2 we saw that there was no difficulty to derive the basic
equation for the multiplier.  It is the same equation as for the
determination of a multiplier for a one-degree-of-freedom system.
In the context of our presentation the main implication of that
result was that Lie's method was equally applicable to systems of
more than one degree of freedom.  For the sake of simplicity of
presentation we confined our attention to just two dependent
variables, but it is quite obvious that the result is general.
When we considered the specific example of the system, (\ref
{3.1}/\ref {3.2}), we used obvious solutions of (\ref {2.7}) to
obtain a class of Lagrangians which includes the usual Lagrangian
for such a system.  Furthermore we demonstrated the use of the
symmetries of the system to construct two other multipliers and
hence a second integral of the system \footnote {Obviously the
Hamiltonian corresponding to the Lagrangian (\ref {3.11}) with the
gauge function set at zero gives one integral.}. Consequently the
system, (\ref {3.1}/\ref {3.2}), is integrable in the sense of
Liouville.  This is to be expected given that the system is a pair
of coupled linear oscillators.  In addition to the `standard'
Lagrangian\footnote{The Lagrangian (\ref{3.11}) can be obtained
from the matrix $C_{2567}$ which has a constant determinant.},
(\ref {3.11}), we explicitly demonstrated the existence of another
Lagrangian. One infers the existence of many.

\strut\hfill

 It is amazing to think that in the case of a two-dimensional
linear oscillator, which admits a 15-dimensional Lie point symmetry
algebra (the maximum for a two-dimensional system),  the possible Lagrangians have to be sought among 1365
different determinants!

\strut\hfill

Further work is in progress in order to address the case of
dissipative systems.

\section*{Acknowledgments} This work was undertaken while PGLL was
enjoying the hospitality of Professor MC Nucci and the facilities of
the Dipartimento di Matematica e Informatica, Universit\`a di
Perugia. The continued support of the University of KwaZulu-Natal is
gratefully acknowledged.

\end {document}